\documentstyle[11pt,aaspp4,epsfig,graphicx]{article}
\begin{document}

\def\simg{\mathrel{\hbox{\rlap{\lower.55ex \hbox {$\sim$}}
                   \kern-.3em \raise.4ex \hbox{$>$}}}}
\def\siml{\mathrel{\hbox{\rlap{\lower.55ex \hbox {$\sim$}}
                   \kern-.3em \raise.4ex \hbox{$<$}}}}
\hfill{\it Revised, ApJ subm 3/15/01 (7/27/00)}

\title{A POSSIBLE INTRINSIC FLUENCE-DURATION \\
 POWER-LAW RELATION IN GAMMA-RAY BURSTS}

\author{L.G. Bal\'azs$^1$, P. M\'esz\'aros$^{2,3}$, Z.  Bagoly$^4$, 
I. Horv\'ath$^5$ \& A. M\'esz\'aros$^6$ }


\bigskip\noindent
$^1$Konkoly Observatory, Budapest, Box 67, H-1525,
Hungary; balazs@ogyalla.konkoly.hu\\
$^2$Dept. of Astronomy \& Astrophysics, Pennsylvania State
University, 525 Davey Lab. University Park, PA 16802,
USA; nnp@astro.psu.edu\\
$^3$Institute of Astronomy, University of Cambridge, Madingley Road,
 Cambridge CB3 0HA, England, U.K.\\
$^4$Laboratory for Information Technology, E\"{o}tv\"{o}s University,
  P\'azm\'any P\'eter s\'et\'any 1/A, H-1518,
Hungary; bagoly@ludens.elte.hu\\
$^5$Department of Physics, ZMNE BJKMFK, Budapest, Box 12, H-1456,
Hungary; hoi@bjkmf.hu\\
$^6$Department of Astronomy, Charles University, 180 00 Prague 8,
 V Hole\v sovi\v ck\'ach 2, Czech Republic; meszaros@mbox.cesnet.cz\\

\begin{abstract}

We argue that the distributions of both the intrinsic fluence and the intrinsic
duration of the $\gamma$-ray emission in gamma-ray bursts from the BATSE sample
are well represented by log-normal distributions, in which the intrinsic dispersion 
is much larger than the cosmological time dilatation and redshift effects. 
We perform separate bivariate log-normal distribution fits to the BATSE short and 
long burst samples. The bivariate log-normal behavior results in an ellipsoidal 
distribution, whose major axis determines an overall statistical relation between 
the fluence and the duration. We show that this fit provides evidence for a 
power-law dependence between the fluence and the duration, with a statistically 
significant different index for the long and short groups. We discuss possible 
biases  which might affect this result, and argue that the effect is probably real. 
This may provide a potentially useful constraint for models of long and short bursts.

\end{abstract}


\section{Introduction}
\label{sec:intro}

The simplest grouping of gamma-ray bursts (GRBs), which is still lacking
a clear physical interpretation, is given by their well-known bimodal
duration distribution. This divides bursts into long ($T \simg 2$ s)
and short ($T \siml 2$ s) duration groups (\cite{k}), defined through
some specific duration definition such as $T_{90}$, $T_{50}$ or similar.
The bursts measured with the BATSE instrument on the Compton Gamma-Ray
Observatory are usually characterized by 9 observational quantities, i.e. 
2 durations, 4 fluences and 3 peak fluxes (\cite{me96,pac,me00}).
In a previous paper (\cite{bag98}), we used the principal components 
analysis (PCA) technique to show that these 9 quantities
can be reduced to only two significant independent
variables, or principal components (PCs).
These PCs can be interpreted as principal vectors, which are
made up of some subset of the original observational quantities.
The most important PC is made up essentially by the durations 
and the fluences, while the second, weaker PC is largely made up of 
the peak fluxes. This simple observational fact, that the dominant 
principal component consists mainly of the durations and the fluences,
may be of consequence for the physical modeling of the burst mechanism.

In this paper we investigate in greater depth the nature of
this principal component decomposition, and in particular, we analyze 
quantitatively the relationship between the fluences and durations 
implied by the first PC.  In our previous PCA treatment of the BATSE 
Catalog \cite{pac} we used logarithmic variables, since these are useful 
for dealing with the wide dynamic ranges involved. Since the logarithms 
of the durations and the fluences can be explained by only one quantity 
(the first PC), one might suspect the existence of only one physical 
variable responsible for both of these observed quantities. The PCA 
assumes a linear relationship between the observed quantities and the 
PC variables. The fact that the logarithmic durations and fluences can 
be adequately described by only one PC implies a proportionality between 
them and, consequently, a power law relation between the observed 
durations and fluences.

We analyze the distribution of the observed fluences and durations of 
the long and the short bursts, and we present arguments indicating that
the intrinsic durations and fluences are well represented by log-normal 
distributions. The implied bivariate log-normal distribution 
represents an ellipsoid in these two variables, whose major axis 
inclinations are statistically different for the long and the short bursts.
An analysis of the possible biases and complications is made,
leading to the conclusion that the relationship between the 
durations and fluences appears to be intrinsic, and may thus be
related to the physical properties of the sources themselves.
We calculate the exponent in the power-laws for the two types 
of bursts, and find that for the short bursts the energy fluence is 
roughly proportional to the intrinsic duration, while for the long ones 
the fluences are roughly proportional to the square of intrinsic durations. 
The possible implications for GRB models are briefly discussed.

\section{Analysis of the Duration Distribution}
\label{sec:durations}

Our GRB sample is selected from the current BATSE Gamma-Ray Burst 
Catalog according to two criteria, namely, that they have both
measured $T_{90}$ durations and fluences (for the 
definition of these quantities see \cite{me00}, henceforth 
referred to as the Catalog). The Catalog in its final version 
lists 2041 bursts for which a value of $T_{90}$ is given.
The fluences are given in four different energy channels,
$F_1, F_2, F_3, F_4$, whose energy bands correspond to
$[25,50]$ keV, $[50,100]$ keV, $[100,300]$ keV and
$> 300 $ keV. The ``total" fluence is defined as 
$F_{tot} = F_1 + F_2 + F_3 + F_4$, and we restrict our sample to 
include only those GRBs which have $F_i>0$ values in at least 
the channels $F_1, F_2, F_3$. Concerning the fourth channel, whose 
energy band is $>300$ keV, if we had required $F_4>0$ as well this
would have reduced the number of eligible GRBs by $\simeq 20\%$. Hence, 
we decided to accept also the bursts with $F_4 = 0$, rather than deleting 
them from the sample. (With this choice we also keep in the sample the 
no-high-energy (NHE) subgroup defined by \cite{pen97}.)
Our choice of $F\equiv F_{tot}$, instead of some other quantity as the main
variable, is motivated by two arguments. First, as discussed in \cite{bag98},
$F_{tot}$ is the main constituent of one of the two PCs which represent the
data embodied in the BATSE catalog, and hence it can be considered as a 
primary quantity, rather than some other combination or subset of its
constituents. Second, Petrosian and collaborators in a series of articles 
(\cite{ep92}, \cite{pl96}, \cite{lp96}, \cite{lp97}) have also argued
for the use of the fluence as the primary quantity instead of, e.g., the
peak flux. 
Using therefore these two cuts, we are left with $N=1929$ GRBs, all of 
which have defined $T_{90}$ and $F_{tot}$, as well as peak fluxes $P_{256}$.
This is the sample that we study in this paper.

The distribution of the logarithm of the observed $T_{90}$ displays two 
prominent peaks\footnote{There is also evidence for the existence of a
third subgroup as part of the long duration group, see \cite{hor}, \cite{muk}, 
\cite{hak00a}, \cite{HA}, which shows a distinct sky angular distribution 
(M\'esz\'aros, 2000a,2000b). We do not deal with this third group here.},
which is interpreted as reflecting the existence of two groups of GRBs (\cite{k}).
This bimodal distribution can be well fitted by means of two Gaussian
distributions (e.g., \cite{hor}), indicating that both the long and 
the short bursts are individually well fitted by pure Gaussian 
distributions in the logarithmic durations.

The fact that the distribution of the BATSE $T_{90}$ quantities within a group 
is log-normal is of interest, since we can show that this property may be 
extended to the intrinsic durations as well. 
Let us denote the observed duration of a GRB with $T_{90}$ (which may be 
subject to cosmological time dilatation), and denote with $t_{90}$ 
the duration which would be measured by a comoving observer, i.e. the
intrinsic duration. One has then
\begin{equation}
 T_{90} = t_{90} f(z)
\label{eq:t90}
\end{equation}
where $z$ is the redshift, and $f(z)$ measures the time dilatation.
For the concrete form of $f(z)$ one can take $f(z) = (1+z)^k$,
where $k=1$ or $k=0.6$, depending on whether energy stretching is included
or not (see \cite{fb95} and \cite{mm96}). If energy stretching is included, 
for different photon frequencies $\nu$ the $t_{90}$ depends on these 
frequencies as $t_{90}(\nu) = t_{90}(\nu_o) (\nu/\nu_o)^{-0.4} \propto 
\nu^{-0.4}$, where $\nu_o$ is an arbitrary frequency in the measured range 
(i.e. for higher frequencies the intrinsic duration is shorter). 
The observed duration at $\nu$ is simply $(1+z)$ times the intrinsic
duration at $\nu \times (1+z)$. Thus, $T_{90}(\nu) = t_{90}(\nu (1+z)) (1+z)$ =
$t_{90}(\nu_o)(\nu (1+z)/\nu_o)^{-0.4} (1+z) = t_{90}(\nu)(1+z)^{0.6}$.
Hence, when stretching is included, $f(z) = (1+z)^{0.6}$ is used. 

Taking the logarithms of both sides of equation (\ref{eq:t90})
one obtains the logarithmic duration as a sum of two independent
stochastic variables. According to a theorem of \cite{cra37} (see also
\cite{renyi}), if a variable $\zeta$ which has a Gaussian distribution is given 
by the sum of two independent variables, e.g. $\zeta = \xi + \eta$, then both
$\xi$ and $\eta$ have Gaussian distributions. Therefore, the Gaussian 
distribution of $\log T_{90}$ (confirmed for the long and short groups 
separately, \cite{hor}) implies that the same type of distribution exists 
for the variables $\log t_{90}$ and $\log f(z)$. 
However, unless the space-time geometry has a very particular structure, 
the distribution of $\log f(z)$ cannot be Gaussian.  This means that the 
Gaussian nature of the distribution of $\log T_{90}$ must be dominated 
by the distribution of $\log t_{90}$, and therefore the latter must then 
necessarily have a Gaussian distribution. 
This holds for both duration groups separately. 
This also implies that the cosmological time dilatation should
not affect significantly the observed distribution of $T_{90}$, 
which therefore is not expected to differ statistically from 
that of $t_{90}$. (We note that several other authors, e.g. \cite{wp94}, 
\cite{nor94}, \cite{nor95}, have already suggested that the distribution 
of $T_{90}$ reflects predominantly the distribution of $t_{90}$.)

One can check the above statement quantitatively by calculating the
standard deviation of $f(z)$, using the available observed redshifts of
GRB optical afterglows. The number of the latter is, however, 
relatively modest, and so far they have been obtained only for long 
bursts. There are currently upwards of 18 GRBs with known redshifts 
(\cite{cg}, \cite{bloom01}, \cite{bloom01b}).
The calculated standard deviation is $\sigma_{\log f(z)}=0.14$,
assuming $\log f(z)=\log (1+z)$.  Comparing the variance $\sigma^2_{\log f(z)}$ 
with that of the group of long burst durations (see Table 1, which gives
$\sigma_{\log T_{90}} = 0.37$), one infers that the standard deviation of 
$\log f(z)$, or $\log (1+z)$, can explain only about 
$(0.14/0.37)^2 \simeq 14\%$ of the total variance of the 
logarithmic durations. (If $ f(z) = (1+z)^{0.6}$, then the standard 
deviation of $\log f(z)$ can only explain an even smaller
amount than 14\%, because $\sigma_{\log f(z)} = 0.6\times 0.14$.)
This comparison gives support to the conclusion obtained by
applying Cram\'er's theorem to the long duration group.

\section{Distribution of the Energy Fluences}
\label{sec:energies}

The observed total fluence $F_{tot}$ can be expressed as
\begin{equation}
F_{tot} = \frac{(1+z)E_{tot}}{4\pi d_l^2(z)} = c(z) E_{tot}.
\label{eq:fluence}
\end{equation}
Here $E_{tot}$ is the total emitted energy of the GRB at the source
in ergs, the total fluence has dimension of erg/cm$^2$, and $d_l(z)$ is the 
luminosity distance corresponding to $z$ for which analytical expressions 
exist in any given Friedmann model (e.g. \cite{weinberg72}, \cite{peebles93}). 
(We note that the considerations in this paper are valid for any Friedmann model.
Note also that the usual relation between the luminosity and flux is
given by a similar equation without the extra $(1+z)$ term in the
numerator. Here this extra term is needed because both the left-hand-side
is integrated over observer-frame time an the right-hand-side is
integrated over time at the source).

Assuming as the null hypothesis that the $\log F_{tot}$ of the 
short bursts has a Gaussian distribution, for the sample of 
447 bursts with $T_{90} < 2$ s, a $\chi^2$ test with 26 degrees 
of freedom gives an excellent fit with $\chi^2=20.17$.
Accepting the hypothesis of a Gaussian distribution within
this group, one can apply again Cramer's theorem similarly to what
was done for the logarithm of durations. This leads to the conclusion 
that either {\it both} the distribution of $\log c(z)$ and the 
distribution of $\log E_{tot}$ are Gaussian, or else the variance of 
one of these quantities is negligible compared to the other, which then
must be mainly responsible for the Gaussian behavior.

The above argument, however, should be taken with some caution. 
As shown in \cite{bag98}, the stochastic variable corresponding to 
the duration is independent from that of the peak flux.
This means that a fixed level of detection, given by the peak fluxes, 
does not have significant influence on the shape of the detected 
distribution of the durations 
(e.g. \cite{ep92}, \cite{wp94}, \cite{nor94}, \cite{nor95},
\cite{pl96}, \cite{lp96}, \cite{lp97}). 
In the case of the fluences, however, a detection threshold in the peak
fluxes induces a bias on the true distribution, since they are 
stochastically not independent. Therefore the log-normal distribution 
recognized in the data does not necessarily imply the same behavior 
for the true distribution of fluences occurring at the detector.
A further complication arises from the differences of the spectral
distribution among GRBs. A discussion of these problems can be 
found in a series of papers published by Petrosian and collaborators
({\cite{ep92}, \cite{pl96}, \cite{lp96}, \cite{lp97}, \cite{lp99}).
Despite these difficulties, there are substantial reasons to
argue that the observed distribution of fluences is dominated by
the intrinsic distribution.
This assumption can be tested by comparing the variance
of the fluences with those obtained for $c(z)$ considering
the GRBS that have measured $z$. We will return to this problem 
in more detail in \S \ref{sec:bias} dealing with the effect of 
possible observational biases.

A Gaussian behavior of $\log c(z)$ can almost certainly be
excluded. One can do this on the basis of the current observed
distribution of redshifts (e.g. \cite{bloom01,bloom01b}), or on the basis of
fits of the number vs. peak flux distributions (e.g.
\cite{fb95}, \cite{uw95}, \cite{hor3}, \cite{rm97}).
In such fits, using a number density $n(z) \propto (1+z)^D$ with
$D \simeq (3-5)$, one finds no evidence for the stopping of this 
increase with increasing $z$ (up to $z \simeq (5-20)$). 
Hence, it would be contrived to deduce from this result that 
the distribution of $\log c(z)$ is normal.
In order to do this, one would need several ad hoc
assumptions. First, the increasing of number density would
need to stop around some unknown high $z$. This was studied, e.g. 
by \cite{mm95}, \cite{hor3}, \cite{mm96}, and no such
effect was found. Second, even if this were the case, above this
$z$ the decrease of $n(z)$ should mimic the behavior of a log-normal
distribution for $c(z)$, without any obvious justification.
Third, below this $z$ one must again have a log-normal behavior
for $c(z)$, in contradiction with the various number vs. peak flux fits.
Fourth, this behavior should occur for any subclass separately.
Hence, the assumption of log-normal distribution of $c(z)$ appears highly
improbable, and this holds for both groups of GRBs.

Thus, for the short bursts the variance of $\log c(z)$ must be
negligible compared to the variance of $\log E_{tot}$.
The latter possibility means that the observed distribution of 
the logarithmic fluences is essentially intrinsic, and therefore
$\log E_{tot}$ should have a Gaussian distribution for the group of 
short bursts.

In the case of the long bursts, a fit to a Gaussian distribution 
of logarithmic fluences does not give a significance level which 
is as convincing as for the short duration group.
For the 1482 GRBs with $T_{90} > 2$ s a $\chi^2$ test on $\log F_{tot}$
with 22 degrees of freedom gives a fit with $\chi^2=35.12$.
Therefore, in this case the $\chi^2$ test gives only a low
probability of 3.5\% for a Gaussian distribution. 
This circumstance prevent us from applying Cram\'er's theorem
directly in the same way as we did with the short duration group.
Calculating the variance of $\log c(z)$ for the GRBs with known redshifts
\cite{bloom01b} one obtains $\sigma_{\log c(z)}=0.43$. From Table 1 
of this article it follows that $\sigma_{\log F_{tot}}=0.66$. 
Hence, the variance of $c(z)$ gives roughly a $(0.43/0.66)^2 \simeq 43\%$ 
contribution to the entire variance, which is also a larger value than in 
the case of the durations. 
We return to this question in \S \ref{sec:correlation}.

\section{Fitting the Logarithmic Fluences and Durations
by the Superposition of two Bivariate Distributions}
\label{sec:bivariate} 

We assume here as a working hypothesis that the distributions of the 
variables $T_{90}$ and $F_{tot}$, for both the short and long groups, 
can be approximated by log-normals. In this case, it is possible 
to fit {\it simultaneously} the values of $\log F_{tot}$ and $\log T_{90}$ 
by a single two-dimensional (bivariate) normal distribution. This 
distribution has five parameters (two means, two dispersions, and the 
correlation coefficient). Its standard form is 
$$ f(x,y) dx dy =
\frac{N}{2 \pi \sigma_x \sigma_y \sqrt{1-r^2}}\times $$
\begin{equation}
\exp\left[-\frac{1}{2(1-r^2)}\left(\frac{(x-a_x)^2}{\sigma_x^2} +
\frac{(y-a_y)^2}{\sigma_y^2} - \frac{2r(x-a_x)(y-a_y)}
{\sigma_x \sigma_y}\right)\right]
 dxdy\;,
\label{eq:exp}
\end{equation}
where $x =\log T_{90}$, $y = \log F_{tot}$ $a_x$, $a_y$ are the means,
$\sigma_x$, $\sigma_y$ are the dispersions, and $r$ is the correlation
coefficient (\cite{tw53}; Chapt. 1.25).
An equivalent set of parameters consists of taking
the same two means with two other dispersions $\sigma_{x}^{,}$
$\sigma_{y}^{,}$, and (instead of the correlation coefficient)
the angle $\alpha$ between the axis $\log T_{90}$ and the semi-major
axis of the ``dispersion ellipse". (In the case of bivariate normal
distributions, the constant probability curves define ellipses with
well-defined axis directions). In this case $\alpha$ and
the correlation coefficient are related unambiguously
through the analytical formula
\begin{equation}
\tan 2 \alpha = \frac{2r\sigma_x \sigma_y}{\sigma_y^2 - \sigma_x^2}\;,
\label{eq:tan}
\end{equation}
and the relations among the variances are also given by analytical
formulae (\cite{tw53}; Chapt. 1.26).
If the data are well fitted by this bivariate normal distribution, then
the distributions of each of the variables by themselves must also be 
univariate normal distributions (the marginal distributions are also normal).

A crucial point in this analysis is that, when the 
$r$-correlation coefficient differs from zero, then
the semi-major axis of the dispersion ellipse represents a linear
relationship between $\log T_{90}$ and $\log F_{tot}$, with a slope 
of $m=\tan \alpha$. This linear relationship between the logarithmic 
variables implies a power law relation of form $F_{tot} = (T_{90})^m$ 
between the fluence and the duration, where $m$ may be different for 
the two groups. As we have shown in \S2. a similar relation 
will exist between $t_{90}$ and $E_{tot}$.

Fitting the data with the superposition of two bivariate 
log-normal distributions can be done by a standard search for 
11 parameters with $N=1929$ measured points (cf. \cite{pre}; Chapt. 15). 
(Both log-normal distributions have five parameters; the eleventh parameter 
defines the weight of the first log-normal distribution.) We will use 
$\tan \alpha$ as the fifth parameter for both partial distributions 
(``terms").  Figure 1. shows the values of $x =\log T_{90}$ and 
$y =\log F_{tot}$ for the $N= 1929$ GRBs.
Each GRB defines a point in the $x, y$ plane
with coordinates $x_i, y_i$ $(i = 1,2,...,N)$. The fitted function
$f_2(x,y, 
a_{xk}, a_{yk},\sigma_{xk}^{,}, \sigma_{yk}^{,}, \alpha_k, W)$
($k=1,2$) is a sum of two normal distributions as given in
Eq.(\ref{eq:exp}).
The normalization constant of the first [second] term is $NW$ [$N(1-W)$],
where $W$ is the weight ($0 \leq W \leq 1$). For the first (second) term
the parameters are $a_{x1}, a_{y1},\sigma_{x1}^{,}, \sigma_{y1}^{,}, 
\alpha_1$ ($a_{x2}, a_{y2},\sigma_{x2}^{,}, \sigma_{y2}^{,}, \alpha_2$).

We obtain the best fit to the 11 parameters through a maximum likelihood
(ML) estimation (e.g., \cite{KS76}, Vol.2., pp.57-58).
We search for the maximum of the formula
\begin{equation}
L_2 = \sum_{i=1}^{N} \ln f_2(x_i, y_i,
a_{xk}, a_{yk},\sigma_{xk}^{,}, \sigma_{yk}^{,}, \alpha_k, W)
\label{eq:ml}
\end{equation}
using a simplex numerical procedure (\cite{pre},
Chapt. 10.4); the index ``2" in $L_2$ shows that we have a sum of two
log-normal distributions of type given by Eq.(\ref{eq:exp}).
The results of this fit are shown in Table 1. We discuss this fit
in the next \S, together with a discussion of instrumental biases.

\section{Bias Effects and Fit Results}
\label{sec:bias}

Several papers discuss the biases in the BATSE values of
$F_{tot}$ and $T_{90}$ (cf. \cite{ep92}, \cite{lamb93}, \cite{lp96}, 
\cite{pl96}, \cite{lp97}, \cite{pac}, \cite{hak00}, \cite{meg00}). 
The effect of these biases is non-negligible, and they may in principle
have an impact on the correlations between fluence and duration.
In other words, it could be that the correlations between the measured 
fluences and measured durations do not necessarily reflect
(due to several instrumental effects) the actual correlations between 
the real fluences and durations (i.e. between the ideal data which 
would be obtained by ideal bias-free instruments). In this Section
we discuss several tests which indicate that these biases do not 
significantly influence the final results presented. 

Roughly speaking, there are two kinds of instrumental biases. First, some 
of the faint GRB below the detection threshold may not be detected, and 
these 
missing GRBs - if they were detected - could change (at least in principle)
the statistical relations between the durations and fluences.
Second, even for GRB which are detected, the measured $F_{tot}$ and 
$T_{90}$ do not reproduce the real values of these quantities,
due to the different background noise effects and other complications 
mainly in the values of $F_{tot}$ (for a more detailed survey of biases 
see the review of \cite{meg00}).

Both of these types of biases are particularly important for the
fainter GRBs. To evaluate the impact of these effects, without going 
into instrumentation details, we will perform two different sets of
test calculations. First, we will truncate the whole sample of GRBs 
with respect to the peak flux, and we will restrict ourselves to the 
brighter ones. For a sufficiently high truncation limit, this truncated 
sample should be free from the effects of the first type of bias above.
Second, we will modify the measured values of $F_{tot}$ and $T_{90}$ 
in order to approximate them by real bias-free values.  Then we repeat 
the calculations of the previous Section for these test samples.

To restrict ourselves to the brighter GRBs we do two truncations.
First, we take only GRBs with $P_{256} > 0.65$ photon/$(cm^2 s)$ 
(N = 1571); second, we take only GRBs with $P_{256} > 1.26$ 
photon/$(cm^2 s)$ (N= 994).  The first choice is motivated by the
analysis of \cite{pen97}, and this should already cancel the impact 
of biases of first type (\cite{stern}, \cite{pen97}). The second choice 
is an ad-hoc one, and is motivated by two opposite requirements. Clearly, 
to avoid the impact of biases of second type, the truncation should be 
done on the highest possible peak flux value. On the other hand, the 
number of remaining bright GRBs should not be small compared with the 
whole sample.  A sample with $P_{256} > 1.26$  photon/$(cm^2 s)$ (N= 994)
appears to be a reasonable choice from these opposite points of view; 
for these high peak fluxes the biases in the values of $F_{tot}$ and 
$T_{90}$ should be largely negligible.

The results after the first truncation are collected in Table 2 and 
can be seen in Figure 2. We see that all the values are practically 
identical with the values of Table 1. To complete the ML estimation,
we need to calculate also the uncertainties in the obtained best fit 
parameters. For our present purposes, it is enough to obtain these 
uncertainties for $\alpha_1$ and $\alpha_2$, respectively.
For this we use the fact that
\begin{equation}
2 ( L_{2,max} -  L_2) \simeq \chi^2_{11},
\label{eq:chi}
\end{equation}
where $L_{2,max}$ is the likelihood for the best parameters,
and $L_2$ is the likelihood for the true values of the parameters,
cf. \cite{KS76}. The $\chi^2_{11}$ is the value of $\chi^2$
function for 11 degrees of freedom (with the degrees of freedom given
by the number of parameters). Taking the values of $\chi^2_{11}$
corresponding to $1\sigma$ probability ($\simeq 68\%$) one obtains a 
10 dimensional hypersurface in the 11 dimensional parameter-space 
around the point defined by the best fit parameters. This hypersurface
can be obtained through computer simulations by changing the values 
of the parameters around the best fit values and for any new set 
calculating the value of the likelihood. Then,
using Eq.(\ref{eq:chi}), one can estimate the $1 \sigma$ uncertainty
in the parameters. This procedure leads to the values
$0.87 \leq \tan \alpha_1 \leq 1.21$
and $2.02 \leq \tan \alpha_2 \leq 2.52$, respectively.
In other words, there is a 68\% probability that the tangents of angles
are in these intervals.

The random probability of obtaining identical angles for both groups 
may be calculated as follows. We do again an ML estimation, similarly 
to the procedure used to derive Table 2, except for one difference. 
That is, we consider only 10 independent parameters, because we require 
that the two angles be identical. Doing this, we obtain that the 
$L_{2,max}$ value is smaller by $4.46$ than the value obtained 
for 11 parameters. (The concrete values of the 10 parameters are 
unimportant here.) Then the one degree of freedom $\chi^2 = 8.92$, 
corresponding to the difference between the 11 and 10 parameter ML 
estimation (see Eq.(\ref{eq:chi})) defines a 0.3\% probability. 
This is the random probability that the two angles are identical 
(\cite{KS76}).  The two angles are therefore definitely different, 
with a high level of significance.

Thus, for the two groups the dependence of the total fluence on the 
measured duration is 
\begin{equation}
F_{tot} \propto \cases{ (T_{90})^{1}~~;&~~(short bursts); \cr
                        (T_{90})^{2.3}~~;&~~(long bursts) \cr}~,
\label{eq:ftotpower}
\end{equation}
and the exponents of the two groups are different at a high level of
confidence, with $\siml 0.3\%$ probability of random occurrence.

The results obtained with the second truncation are collected in Table 3. 
We see that all values are again similar to the values of Table 1 and 
Table 2. The difference in the averages of the durations and fluences
understandable, because we take brighter GRBs.  However the changes in 
them are small, and our conclusions therefore remain the same.

To further verify this result, we also perform a second type of test.
We modify the values of the measured $F_{tot}$ and $T_{90}$ in order 
to approximate them by their real bias-free values. We consider a 
very simplified model of GRB pulse shapes, namely, we assume that 
the measured time behavior of a GRB may be described by a triangle.
If a GRB pulse arises at a time instant $t_1$, after this time its
flux increases linearly, and reaches its peak flux (denoted by $P$)
at time $t_2$. Then it again decreases linearly, and the flux becomes
zero at a time instant $t_3$. Then the real measured duration is 
$t = t_3 - t_1$, and the real fluence is $F = t \times P/2$. 
Assume now that there is a background noise, causing a flux $P_o$. 
Then, clearly, the GRB will be detected only when the flux $P$
is larger than $P_o$. This means that the measured duration will be 
$t' = (1- P_o/P)t$. Also, the measured fluence will be lower, because 
no flux is seen during the time when the flux is smaller than $P_o$. 
The value of the measured fluence will be given in this case by
$F' = (1 - (P_o/P)^2) F$. Taking this into account, we will do the 
modifications $F_{tot;mod} = F_{tot}/(1 - (P_o/P_{256})^2)$ and 
$T_{90;mod} = T_{90}/(1 - P_o/P_{256})$, respectively. This means 
that the modified values (which are expected to be closer to the real 
ones) are larger than the measured ones. There are of course possible
objections that can be raised against such a procedure. One might argue 
that these modifications are ad-hoc for several reasons, e.g., that
$t'$ and $t$ in the previous consideration are not identical to $T_{90}$ 
and $T_{90;mod}$; or similarly, that $F$ and $F'$ are not identical to 
$F_{tot}$ and $T_{tot;mod}$; also that $P$ cannot be substituted 
automatically by $P_{256}$; that the concrete value of $P_o$ is subject 
to change; that the "triangle" approximation is arbitrary; etc. 
Nevertheless, keeping all these caveats in mind, it is still useful to 
see what would be the change, if the calculation of the previous Section 
is repeated for these modified fluences and durations.

The results of this second test calculation are collected in Table 4. 
The concrete value of $P_o = 0.3$ photon/(cm$^2$s) is based on the results
of \cite{stern} and \cite{pen97}, which suggest that the background noise 
is $\simeq (0.2 - 0.3)$ photon/(cm$^2$s). In order to avoid the problem 
with GRBs which have $P_{256} < P_o$, here we do not use the whole sample 
of $N=1929$, but a sample with $P_{256} > 0.65$ photon/(cm$^2$s) 
($N= 1571$). This truncation does not lead to any essential changes, 
as was already seen earlier in this Section.  The values of Table 4 are 
again very similar to those of Table 2. Omitting the calculation of 
the uncertainties in the best parameters, we calculate only the
probability of having identical angles; this probability is 0.4\%
(the value of $\chi^2$ drops by $4.22$). This tends to support, again, 
the earlier results obtained with the larger sample.

Therefore, from these two different tests we are led to conclude that 
the instrumental biases do not change the basic results.
The discussion in this Section gives support to the interpretation that 
the correlation between the logarithms of the durations and the logarithms 
of the fluences are real, that they are different for the long and the
short bursts, and that these conclusions remain valid even after taking
into account instrumental biases.

\section{Are the Correlations Actually Intrinsic?}
\label{sec:correlation}

In the previous Section we presented arguments showing that the 
different power law relations between $F_{tot}$ and $T_{90}$, expressed
through Eq.(\ref{eq:ftotpower}), are real, and are not substantially 
affected by instrumental bias effects. In \S \ref{sec:energies} it was
shown that these same power-law 
relations hold between the intrinsic $t_{90}$ and $E_{tot}$ values. 
Since, however, for the long bursts the validity of a log-normal
representation of the fluences is not so obvious, we return here
for the sake of completeness to this question. 

In general, the PCA analysis shows that the logarithm of any measured 
quantity can be represented by a linear combination of PCs.
In the PCA analysis of \cite{bag98} the whole BATSE sample of GRBs 
(lumping together the long and short groups) it was shown that there 
are two important PCs, which may be identified - to a high accuracy - 
with the $\log T_{90}$ and $\log P_{256}$ variables. This means that 
also the logarithm of fluence may be written as
\begin{equation}
\log F_{tot} = a_1 \log T_{90} + a_2 \log P_{256} + e,
\label{eq:pc}
\end{equation}
where $a_1$ and $a_2$ are some constants (defining the importance of
the PC) and $e$ is some noise term (see \cite{bag98}).

As shown in \S \ref{sec:durations}, the distribution of the measured 
$T_{90}$ is well be described by the superposition of two log-normal 
distributions. It is also shown in \S \ref{sec:durations} that the 
intrinsic durations $t_{90}$ for the two groups (short and long) 
separately are distributed log-normally. Hence, if it were the case
that $a_2$ were negligibly small with respect to $a_1$, it would
be possible to conclude (i.e. without any further separate investigation 
of $\log F_{tot}$ itself) that $\log F_{tot}$ is given by the 
superposition of two log-normal distributions. However, the smallness 
of $a_2$ is not fulfilled generally. Therefore, an additional study of 
$F_{tot}$ is needed, and this is done in \S \ref{sec:energies}. 
It is shown there that, because $c(z)$ in \ref{eq:fluence} is very unlikely
to obey a log-normal distribution (which holds for both groups separately), 
from the log-normal behavior of $\log F_{tot}$ it follows that $E_{tot}$ 
must also have log-normal distribution. More precisely, the log-normal 
distribution of $E_{tot}$ is well justified for the short group, but not 
so well for the long group. Therefore, a further analysis, based on the
PCA method, may help to clarify the situation.

As mentioned, the coefficient $a_2$ in equation (\ref{eq:pc}) is not
always negligible with respect to $a_1$. Nonetheless, a simple trick
can be used which gets one around this obstacle. If we take a narrow 
enough interval of $\log P_{256}$ so that in it it is approximately
valid to take $P_{256} = const$, then $a_2 \log P_{256}$ is also 
constant there, and will not play a role in the form of the 
distribution of $\log F_{tot}$. 
Keeping in mind this possibility, we again do truncations in $P_{256}$
(as in the previous Section), but here we restrict $P_{256}$ from both 
sides (i.e. there is also an upper limit). In addition, we need to take
as narrow an interval of $\log P_{256}$ as possible; of course, the 
number of GRBs remaining cannot be too small (we will require at least 
$N >500$), otherwise a statistical study is not possible. 
Because PCA studies were done for the whole sample, we also do a fitting 
for the whole sample, similarly to what was done in previous Sections.

Results of the truncation with $0.65 < P_{256} < 1.26$ (where $P_{256}$ has
units of photons/(cm s$^2$)) is given in Table 5. We see that the results
are again similar to the whole sample presented in Table 1; of course, 
some concrete values may differ from that of Table 1. due to the
choice of sample. For our purpose {\it here} it
is the most essential conclusion
that the fitting with the superposition of two two-dimensional
log-normal distributions 
may again be done well. This means that the distribution in $\log F_{tot}$
is therefore also a sum of two log-normal distributions.

The results using a truncation with $1.26 < P_{256} < 3.98$ is given in 
Table 6. We again see that the results are similar to those of the whole 
sample given in Table 1. For our purposes here, it is again essential that the 
data is well fitted with a superposition of two two-dimensional log-normal 
distributions; the distribution in $\log F_{tot}$ is therefore also a 
sum of two log-normal distributions.

Having thus concluded that the distribution of $F_{tot}$ is a sum of 
two log-normal distributions, since $c(z)$ is very unlikely to be
log-normally distributed (for both groups separately), $E_{tot}$ should 
also be distributed log-normally for both subclasses separately.

It is appropriate to discuss, in addition, how these results relate
to those showing a cosmological time dilatation and redshift signature.
If it were the case that the correlation were dominantly caused by a 
factor depending on $z$, then the GRBs with lower fluence would have 
systematically longer durations. \cite{nor94,nor95,lp97} find an 
inverse trend between the peak flux $P_{256}$ and the values of 
$T_{50}$ of long bursts $T_{50} >1.5$s, and note that the dispersion 
is substantially larger than the inferred cosmological dilatation effect. 
Our results are not in conflict with that of these authors, although
our purpose is different from theirs. In the present work we have used 
a more extended burst sample than these authors. Our methodology also
differs from that of Norris et al. in at least three respects: we used 
fluences instead of peak fluxes, since we are interested in total energies; 
we used $T_{90}$ instead of $T_{50}$, since that is more useful for 
separating the short and long bursts; and we did not normalize the noise 
to the same level in all the bursts, nor did we perform selection cuts 
designed to find a cosmological dilatation.
As mentioned, in the distribution of the $T_{90}$ we find that the 
dispersion is dominantly intrinsic, which agrees with \cite{nor94,nor95},
and we find that the positive fluence-duration correlation is 
statistically stronger in the short bursts, for which both 
\cite{nor94,nor95} and ourselves find that the dispersion would 
mask a cosmological dilatation signal.  In the Appendix, as a check, 
we compare the behavior of the $T_{50}$ and the $T_{90}$ durations and 
their dispersions as a function of the peak flux. As seen from
Table \ref{tab:a1} of the Appendix, for the long bursts the $T_{50}$ vs.
$P_{256}$ does show (even with our cuts, which are not optimized for that
purpose) a slight trend in the sense of cosmological time dilatation, 
which is substantially weaker than the dispersion. A similar, but 
much fainter trend may be also be seen in the $T_{90}$ vs. $P_{256}$,
which is even more strongly dominated by the dispersion.
For the short bursts (Table \ref{tab:a2} of the Appendix),
not surprisingly, this weak trend cannot be recognized. 
This is in qualitative agreement with the results of \cite{nor94,nor95}. 
We may also ask whether the correlations based on the $T_{90}$
measurements that we discuss here would differ from those derived 
using $T_{50}$. An inspection of Tables \ref{tab:a1} and \ref{tab:a2} 
of the Appendix indicates that the character of the dispersions in both 
variables are essentially the same.  Therefore the correlations and the 
angles (power law indices) would be the same, regardless of whether 
$T_{90}$ or $T_{50}$ is used.

\section{Discussion}
\label{sec:disc}

We have presented evidence indicating that there is a power-law 
relationship between the logarithmic fluences and the $T_{90}$ durations 
of the GRBs in the current BATSE Catalog, based on a maximum likelihood 
estimation of the parameters of the bivariate distribution of these
measured quantities. As shown in the Appendix, the dispersions of the 
$T_{90}$ do not differ significantly from those of the $T_{50}$ 
distributions, and therefore the same correlations and the same power-law 
relations would be expected if one used the $T_{50}$ instead of the 
$T_{90}$. We have also evaluated the possible impact of instrumental 
biases, with the results that the conclusions do not change significantly
when these effects are taken into account.

An intriguing corollary of these results is that the exponents in 
the power-law dependence between fluence and duration differ 
significantly for the two groups of short ($T_{90} < 2$ s) and long 
($T_{90} > 2$ s) bursts. As shown in \S \ref{sec:bias}, this also 
means that the same power law relations hold between the total 
energy emitted ($E_{tot}$) and the intrinsic durations ($t_{90}$) of 
the two groups. The intrinsic nature of this relation is also confirmed 
by further calculations based on a principal component analysis. 

While an understanding of such power-law relations in terms of 
physical models of GRB would require more elaborate considerations, 
without going into details it may be noted here that the results 
are compatible with a simple interpretation where the short bursts 
involve a wind outflow leading to internal 
shocks responsible for the gamma-rays (\cite{rm94,piran99}), in which the 
luminosity is approximately constant over the duration $t$ of the 
outflow, so that both the total energy $E_{tot}$ and the fluence 
$F_{tot}$ are $\propto t$. If an external shock were involved,
e.g. \cite{mr93,piran99}, for a sufficiently short intrinsic duration 
(impulsive approximation) there would be a simple relationship
between the observed duration and the total energy,
$t \propto E^{1/3}$, resulting from the self-similar behavior
of the explosion and the time delay of the pulse arrival from over 
the width of the blast wave from across the light cone. This 
relationship is steeper than the one we deduced for long bursts.
However, the observed $F_{tot} \propto t^{2.3}$ behavior
could possibly be the result of having the observed gamma-rays due 
to some combination of internal and external shocks.

In summary, we have presented quantitative arguments in support of
two new results, namely that there is a power law relation between 
the fluence and duration of GRBs which appears to be physical, and
that this relation is significantly different for the two groups of 
short and long bursts. For the short ones, the total energy released 
is proportional to the duration of the gamma ray emission, while for the 
long ones it is proportional roughly to the square of the duration. 
This may indicate that two different types of 
central engines are at work, or perhaps two different types of progenitor 
systems are involved. It is often argued that those bursts for which 
X-ray, optical and radio afterglows have been found, all of which belong 
to the long-duration group, may be due to the collapse of a massive stellar 
progenitor (e.g. \cite{pa98,fry99}). The short bursts, none of which have 
as of March 2001 yielded afterglows, may be hypothetically associated
with neutron star mergers (e.g. \cite{fry99}) or perhaps other systems.
While the nature of the progenitors remains so far indeterminate, our 
results provide new evidence suggesting an intrinsic difference between 
the long and short bursts, which probably reflects a difference in the 
physical character of the energy release process. This result is completely 
model-independent, and if confirmed, it would provide a potentially useful 
constraint on the types of models used to describe the two groups of bursts.

\bigskip

We are indebted to Dr. G\'abor Tusn\'ady (R\'enyi Institute for Mathematics),
Dr. Chryssa Kouveliotou and Dr. Michael Gibbs (NASA MSFC) and the referee for  
useful discussions and critique. Research supported in part through
OTKA grants T024027 (L.G.B.), F029461 (I.H.) and T034549,
NASA NAG5-2857, Guggenheim Foundation and Sackler Foundation (P.M.), 
GA \v{C}R grant 202/98/0522 and Domus Hungarica Scientiarum et Artium grant (A.M.).


\begin{table}
\caption{The best fit parameters of the sum of two bivariate log-normal
distributions for $x = \log T_{90}$ and $y = \log F_{tot}$ for the
whole sample with $N=1929$, giving a value of likelihood
$\log L_2 = 10 664.82$ (for more details see the text).}
$$
  \begin{array}{rrrrr}
 \hline
   a_{x1}& -0.08       & &   a_{x2} & 1.54     \\
   a_{y1}& -6.22       & &   a_{y2} & -5.29    \\
   \sigma_{x1}^{,}& 0.73   & &  \sigma_{x2}^{,} & 0.67 \\
   \sigma_{y1}^{,}& 0.46   & &  \sigma_{y2}^{,} & 0.37     \\
   \tan \alpha_1 & 0.91& &    \tan \alpha_2 & 2.29    \\
   W & 0.32            & &   &    \\
    \hline
  \end{array}
  $$
\label{tab1}
  \end{table}

\begin{table}
\caption{The best fit parameters of the sum of two bivariate log-normal
distributions for $x = \log T_{90}$ and $y = \log F_{tot}$ for the
sample with $P_{256} > 0.65$
photon/$(cm^2 s)$ ($N = 1571$), giving a value of likelihood
$\log L_2 = 8 388.33$ (for more details see the text).}
$$
  \begin{array}{rrrrr}
 \hline
   a_{x1}& -0.12       & &   a_{x2} & 1.55     \\
   a_{y1}& -6.18       & &   a_{y2} & -5.16    \\
   \sigma_{x1}^{,}& 0.66   & &  \sigma_{x2}^{,} & 0.65 \\
   \sigma_{y1}^{,}& 0.45   & &  \sigma_{y2}^{,} & 0.38     \\
   \tan \alpha_1 & 1.05& &    \tan \alpha_2 & 2.30    \\
   W & 0.32            & &   &    \\
    \hline
  \end{array}
  $$
\label{tab2}
  \end{table}

\begin{table}
\caption{The best fit parameters of the sum of two bivariate log-normal
distributions for $x = \log T_{90}$ and $y = \log F_{tot}$ for the
sample with $P_{256} > 1.26$
photon/$(cm^2 s)$ ($N = 994$), giving a value of likelihood
$\log L_2 = 4906.82$ (for more details see the text).}
$$
  \begin{array}{rrrrr}
 \hline
   a_{x1}& -0.15       & &   a_{x2} & 1.52     \\
   a_{y1}& -5.99       & &   a_{y2} & -4.96    \\
   \sigma_{x1}^{,}& 0.55   & &  \sigma_{x2}^{,} & 0.65 \\
   \sigma_{y1}^{,}& 0.43   & &  \sigma_{y2}^{,} & 0.39     \\
   \tan \alpha_1 & 1.06& &    \tan \alpha_2 & 1.92    \\
   W & 0.30            & &   &    \\
    \hline
  \end{array}
  $$
\label{tab3}
  \end{table}

\begin{table}
\caption{The best fit parameters of the sum of two bivariate log-normal
distributions for the {\it modified}
 $x = \log T_{90}$ and $y = \log F_{tot}$ for the
sample with $P_{256} > 0.65$
photon/$(cm^2 s)$ ($N = 1571$), giving a value of likelihood
$\log L_2 = 8 3512.70$ (for more details see the text).}
$$
  \begin{array}{rrrrr}
 \hline
   a_{x1}& -0.01      & &   a_{x2} & 1.66     \\
   a_{y1}& -6.15       & &   a_{y2} & -5.13    \\
   \sigma_{x1}^{,}& 0.65   & &  \sigma_{x2}^{,} & 0.62 \\
   \sigma_{y1}^{,}& 0.47   & &  \sigma_{y2}^{,} & 0.41     \\
   \tan \alpha_1 & 0.94& &    \tan \alpha_2 & 2.60    \\
   W & 0.33            & &   &    \\
    \hline
  \end{array}
  $$
\label{tab4}
  \end{table}

\begin{table}
\caption{The best fit parameters of the sum of two bivariate log-normal
distributions for
 $x = \log T_{90}$ and $y = \log F_{tot}$ for the
sample with $1.26 >P_{256} > 0.65$
photon/$(cm^2 s)$ ($N = 577$), giving a value of likelihood
$\log L_2 = 2765.64$ (for more details see the text).}
$$
  \begin{array}{rrrrr}
 \hline
   a_{x1}& 0.19      & &   a_{x2} & 1.62     \\
   a_{y1}& -6.44       & &   a_{y2} & -5.44    \\
   \sigma_{x1}^{,}& 0.91   & &  \sigma_{x2}^{,} & 0.45 \\
   \sigma_{y1}^{,}& 0.36   & &  \sigma_{y2}^{,} & 0.22   \\
   \tan \alpha_1 & 0.40& &    \tan \alpha_2 & 1.04    \\
   W & 0.43            & &   &    \\
    \hline
  \end{array}
  $$
\label{tab5}
  \end{table}

\begin{table}
\caption{The best fit parameters of the sum of two bivariate log-normal
distributions for
 $x = \log T_{90}$ and $y = \log F_{tot}$ for the
sample with $3.98 >P_{256} > 1.26$
photon/$(cm^2 s)$ ($N = 667$), giving a value of likelihood
$\log L_2 = 3204.19.64$ (for more details see the text).}
$$
  \begin{array}{rrrrr}
 \hline
   a_{x1}& -0.17      & &   a_{x2} & 1.54     \\
   a_{y1}& -6.15       & &   a_{y2} & -6.15    \\
   \sigma_{x1}^{,}& 0.52   & &  \sigma_{x2}^{,} & 0.36 \\
   \sigma_{y1}^{,}& 0.36   & &  \sigma_{y2}^{,} & 0.52     \\
   \tan \alpha_1 & 0.73& &    \tan \alpha_2 & 1.06    \\
   W & 0.33            & &   &    \\
    \hline
  \end{array}
  $$
\label{tab6}
  \end{table}

\vfill\eject

\begin{figure}[htb]
\figurenum{1}
\plotone{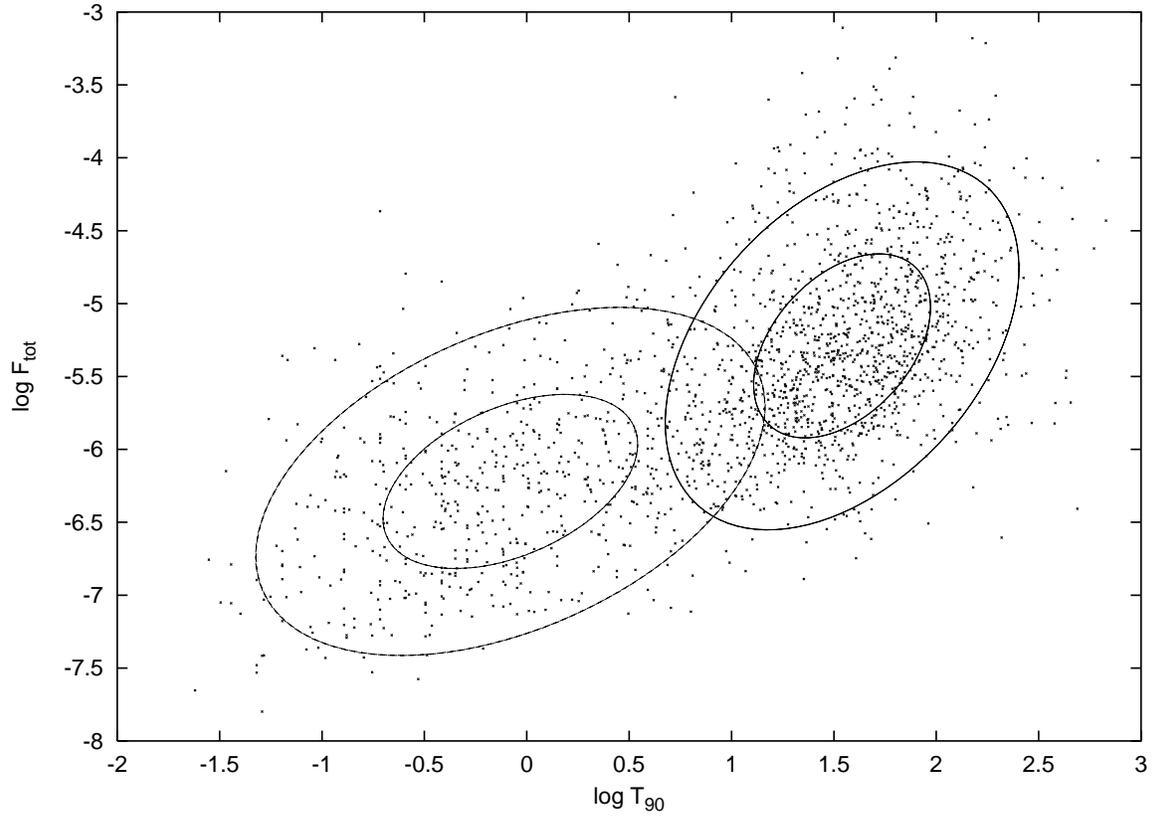}
\caption{The best fit of two bivariate
log-normal distributions for the whole BATSE sample (1929 GRBs).
The ellipses give the $1\sigma$ and
$2\sigma$ probabilities (for more details see the text).}
\end{figure}

\vfill\eject

\begin{figure}[htb]
\figurenum{2}
\plotone{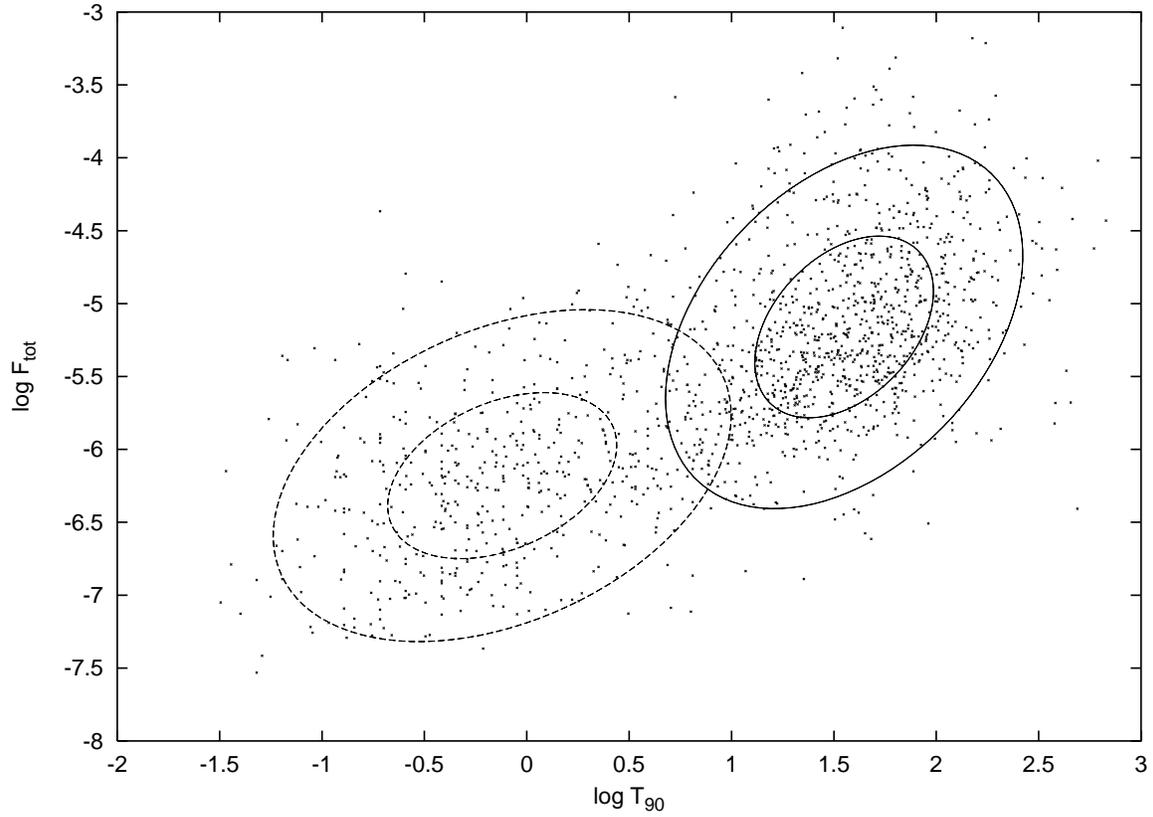}
\caption{The best fit of two bivariate
log-normal distributions for the 1571 GRBs which fulfil the $P_{256} > 0.65 $
photon/$(cm^2 s)$ criterion. The ellipses give the $1\sigma$ and
$2\sigma$ probabilities (for more details see the text).}
\end{figure}

\vfill\eject
\section{APPENDIX: Comparison of $T_{90}$ and $T_{50}$ Statistical
Properties}

In order to check whether there is some influence of the time dilatation on 
the distribution of $T_{90}$ or $T_{50}$ we compare here the basic properties 
of these two quantities in our sample for the long and the short bursts, 
separately. We grouped the data, using the 256 ms peak flux values, into 
0.2 bins in $P_{256}$, and summarized in Tables \ref{tab:a1} and \ref{tab:a2}
the mean values and the corresponding standard deviations of the logarithmic
durations of GRBs in each bin. We stress that this does not include any
equalization of the noise level in the various $p$ bins, and is not intended as a test
of the time dilatation hypothesis, but rather as a test of whether dilatation, 
would have any effect on our results.

Inspecting the durations of long ($T_{90} > 2s$) GRBs summarized in Table 
\ref{tab:a1} one sees that, except from the brightest and faintest bins, 
there is no significant difference in $\log T_{90}$. The decrease of the 
duration in the  faintest bin is probably due to the biasing of the 
determination, namely, the fainter parts of the bursts cannot be discriminated 
against the background and the duration obtained is systematically shorter. 
There is  a remarkable homogeneity and no trend in the standard deviations 
of the $\log T_{90}$.

In the case of the long burst $T_{50}$ durations, this quantity shows an
increasing trend towards bursts of fainter peak flux. The shortening in the
faintest bin is probably also due to selection effects. Similarly to the 
$\log T_{90}$ values the same homogeneity can be observed in the standard 
deviations also in case of $\log T_{50}$. The standard deviations are almost 
the same in both $\log T_{90}$ and $\log T_{50}$.

One can test whether, within our analysis methodology and with our sample,
there is a significant difference among the binned $T_{90}$ values, and 
whether the slight trend  in the $T_{50}$ significantly differs from  zero.
To evaluate the significance of these data we performed a one way analysis 
of variance with the ANOVA program from a standard SPSS package. The ANOVA 
compares the variances within sub-samples of the data (in our case within bins), 
with the variances between the sub-samples (bins).

In the case of $\log T_{90}$  the probability that the difference is 
accidental is 66\%.  In the case of the $T_{50}$ durations the same quantities 
(variances within and between bins) gives a probability of 98.5\% for 
being a real difference between bins, or a probability of 1.5\% that there
is no difference between the bins. This figure gives some significance for 
the reality of a trend in the data; however, this value of 0.2 explains
less than 1/6 of the variance of $T_{50}$ within one bin. We may conclude
that even in this case the variance is mainly intrinsic. 
 
Inspecting the same data in the case of the short duration bursts (Table
\ref{tab:a2}) we come to a similar conclusion, i.e. there is no sign of
trends in the durations of the different bins. Dropping the two faintest 
bins, which are definitely affected by biases, and dropping the poorly 
populated brightest bins, we arrive by the analysis of variances with ANOVA 
to probabilities of 53 \% and 92.1 \% for the difference being purely accidental 
between bins in $T_{90}$ and $T_{50}$, respectively.

\begin{table}
\caption{GRBs of long duration ($T_{90} > 2 s$)}
\begin{tabular}{rrrrrr}
\hline
$\log P_{256}$ & $\log T_{90}$& $\log T_{50}$ & $\sigma_{\log T_{90}}$ &
$\sigma_{\log T_{50}}$ & No. of GRB \\
\hline
    -.50  &   1.24  &    .85 &     .48  &    .47  &    49 \\
    -.30  &   1.42  &   1.00 &     .47  &    .50  &   230 \\
    -.10  &   1.48  &   1.08 &     .49  &    .53  &   309 \\
     .10  &   1.46  &   1.02 &     .51  &    .57  &   272 \\
     .30  &   1.51  &   1.01 &     .52  &    .61  &   194 \\
     .50  &   1.43  &    .94 &     .51  &    .59  &   161 \\
     .70  &   1.45  &    .96 &     .48  &    .56  &   104 \\
     .90  &   1.42  &    .83 &     .54  &    .62  &    56 \\
    1.10  &   1.41  &    .83 &     .50  &    .49  &    44 \\
    1.30  &   1.44  &    .88 &     .50  &    .53  &    34 \\
 $>$1.40  &   1.21  &    .68 &     .41  &    .50  &    29 \\
\hline
\label{tab:a1}
\end{tabular}
\end{table}

\begin{table}
\caption{GRBs of short duration ($T_{90} < 2 s$)}
\begin{tabular}{rrrrrr}
\hline
$\log P_{256}$ & $\log T_{90}$ & $\log T_{50}$ & $\sigma_{\log T_{90}}$ &
$\sigma_{\log T_{50}}$ & No. of GRB \\
\hline
    -.50  &   -.57  &   -.87 &     .55  &    .60  &     7 \\
    -.30  &   -.65  &  -1.01 &     .53  &    .57  &    43 \\
    -.10  &   -.40  &   -.77 &     .49  &    .51  &   103 \\
     .10  &   -.35  &   -.74 &     .35  &    .32  &   105 \\
     .30  &   -.33  &   -.75 &     .39  &    .41  &    75 \\
     .50  &   -.27  &   -.69 &     .35  &    .36  &    54 \\
     .70  &   -.29  &   -.72 &     .36  &    .34  &    25 \\
     .90  &   -.35  &   -.76 &     .39  &    .36  &    22 \\
    1.10  &   -.18  &   -.72 &     .44  &    .39  &     7 \\
    1.30  &   -.74  &  -1.21 &     .31  &    .43  &     5 \\
 $>$1.40  &   -.72  &   -.90 &     .00  &    .00  &     1 \\
\hline
\label{tab:a2}
\end{tabular}
\end{table}


\begin{thebibliography}{}

\bibitem[Bagoly et al. 1998]{bag98} Bagoly, Z., M\'esz\'aros, A.,
Horv\'ath,
I., Bal\'azs, L.G., \& M\'esz\'aros, P. 1998, ApJ, 498, 342.
\bibitem[Bloom et al. 2001]{bloom01} Bloom, J.S., Kulkarni, S.R., \& 
Djorgovski, S.G. 2001, AJ, submitted (astro-ph/0010176)
\bibitem[Bloom et al. 2001b]{bloom01b} Bloom, J.S. and the Caltech GRB group,
 cited in M\'esz\'aros, P. 2001, Science, 291, 79.
\bibitem[Cram\'er 1937]{cra37} Cram\'er, H. 1937, Random variables and
probability distributions, Cambridge Tracts in Mathematics and
Mathematical Physics, No.36 (Cambridge University Press, Cambridge).
\bibitem[Efron \& Petrosian 1992]{ep92} Efron, B., \& Petrosian, V.
1992, ApJ, 339, 345.
\bibitem[Fenimore \& Bloom 1995]{fb95} Fenimore, E.E., \& Bloom,
J.S. 1995, ApJ, 453, 16.
\bibitem[Fryer et al. 1999]{fry99} Fryer, C.L., Woosley, S.E., \&
Hartmann, D. H. 1999, ApJ, 526, 152.
\bibitem[Hakkila et al. 2000a]{hak00a} Hakkila, J., Haglin, D.J., Roiger, R.J., 
Mallozzi, R.S., Pendleton, G.F., \& Meegan, C.A. 2000a, in Gamma-Ray
Bursts; 5th Huntsville Symp., eds. R.M. Kippen, R.S. Mallozzi, G.J.
Fishman, AIP, Melville, p.33.
\bibitem[Hakkila et al. 2000b]{hak00} Hakkila, J., Meegan, C.A., Pendleton,
G.N., Mallozzi, R.S., Haglin, D.J., \& Roiger, R.J. 2000b, in Gamma-Ray
Bursts; 5th Huntsville Symp., eds. R.M. Kippen, R.S. Mallozzi, G.J.
Fishman, AIP, Melville, p.48.
\bibitem[Hakkila et al. 2000c]{HA} Hakkila, J., et al. 2000c.
ApJ, 538, 165.
\bibitem[Horv\'ath 1998]{hor} Horv\'ath, I. 1998, ApJ, 508, 757.
\bibitem[Horv\'ath et al. 1996]{hor3} Horv\'ath, I.,
M\'esz\'aros, P., \& M\'esz\'aros, A. 1996, ApJ, 470, 56.
\bibitem[Kendall \& Stuart 1976]{KS76}
Kendall, M. \& Stuart, A. 1976, The Advanced Theory
of Statistics (Griffin, London).
\bibitem[Kouveliotou et al. 1993]{k}
Kouveliotou, C., et al. 1993, ApJ, 413, L101.
\bibitem[Lamb et al. 1993]{lamb93} Lamb D.Q., Graziani, C., \& Smith, I.A.,
1993, ApJ, 413, L11.
\bibitem[Lee \& Petrosian 1996]{lp96} Lee, T., \& Petrosian, V. 1996, ApJ,
470, 479
\bibitem[Lee \& Petrosian 1997]{lp97} Lee, T., \& Petrosian, V. 1997, ApJ,
474, 37L
\bibitem[Lloyd \& Petrosian 1999]{lp99} Lloyd, N.L., \&
Petrosian, V. 1999, ApJ, 511, 550
\bibitem[Meegan et al. 1996]{me96} Meegan, C.A., et al. 1996, ApJS,
106, 65 (3B BATSE Catalog).
\bibitem[Meegan et al. 2000a]{me00}
Meegan, C.A., et al. 2000a, Current BATSE Gamma-Ray Burst Catalog,
http://gammaray.msfc.nasa.gov/batse/grb/catalog/current/
\bibitem[Meegan et al. 2000b]{meg00}
Meegan C.A., Hakkila, J., Johnson, A., Pendleton, G.,
\& Mallozzi, R.S. 2000b, in Gamma-Ray
Bursts; 5th Huntsville Symp., eds. R.M. Kippen, R.S. Mallozzi, G.J.
Fishman, AIP, Melville, p.43.
\bibitem[M\'esz\'aros \& M\'esz\'aros 1996]{mm96}
M\'esz\'aros, A., \& M\'esz\'aros, P. 1996, ApJ, 466, 29.
\bibitem[M\'esz\'aros et al. 2000b]{mesz00b}
M\'esz\'aros, A., Bagoly, Z., Horv\'ath, I., Bal\'azs, L.G.,
\& Vavrek, R. 2000b, ApJ, 539, 98.
\bibitem[M\'esz\'aros et al. 2000a]{mesz00a}
M\'esz\'aros, A., Bagoly, Z., \& Vavrek, R. 2000a, AA, 354, 1.
\bibitem[M\'esz\'aros \& M\'esz\'aros 1995]{mm95}
M\'esz\'aros, P., \& M\'esz\'aros, A. 1995, ApJ, 449, 9.
\bibitem[M\'esz\'aros \& Rees 1993]{mr93} M\'esz\'aros, P., \& Rees, M. J.
1993, ApJ, 405, 278.
\bibitem[Mukherjee et al. 1998]{muk} Mukherjee, S., et al. 1998,
ApJ, 508, 314.
\bibitem[Norris et al. 1994]{nor94} Norris, J. P., et al. 1994, ApJ, 424,
540.
\bibitem[Norris et al. 1995]{nor95} Norris, J. P., et al. 1995, ApJ, 439,
542.
\bibitem[Norris \& Marani 2000]{cg} Norris, J.P., \& Marani, G. 2000,
http://cossc.gsfc.nasa.gov/batse/counterparts/GRB\_table.html.
\bibitem[Paciesas et al. 1999]{pac} Paciesas, W. S., et al. 1999,
ApJS, 122, 465 (4B BATSE Catalog).
\bibitem[Paczy\'nski 1998]{pa98} Paczy\'nski, B. 1998, ApJ, 497, L45.
\bibitem[Peebles 1993]{peebles93} Peebles, P.J.E. 1993, Principles of
Physical Cosmology (Princeton University Press, Princeton).
\bibitem[Pendleton et al. 1997]{pen97} Pendleton, C.N. 1997, ApJ, 489, 175.
\bibitem[Petrosian \& Lee 1996]{pl96} Petrosian, V. \& Lee, T. 1996, ApJ,
467, 29L.
\bibitem[Piran 1999]{piran99} Piran, T., 1999, Phys.Rep. 314, 575
\bibitem[Press et al. 1992]{pre} Press, W. H.,
Flannery, B.P., Teukolsky, S.A., \&
Vetterling, W.T. 1992, Numerical Recipes (Cambridge University Press,
Cambridge).
\bibitem[Rees \& M\'esz\'aros 1994]{rm94} Rees, M.J., \& M\'esz\'aros,
P. 1994, ApJ, 430, L93.
\bibitem[Reichart \& M\'esz\'aros 1997]{rm97} Reichart, D.E., \&
M\'esz\'aros, P. 1997, ApJ, 483, 597.
\bibitem[R\'enyi 1962]{renyi} R\'enyi, A. 1962. Wahrscheinlichtkeitsrechnung
(VEB Deutscher Verlag der Wissenschaften, Berlin).
\bibitem[Stern et al. 1999]{stern} Stern, B., Poutanen, J., \&
Svensson, R. 1999, ApJ, 510, 312.
\bibitem[Trumpler \& Weaver 1953]{tw53} Trumpler, R. J., \& Weaver, H. F.
1953, Statistical Astronomy (University of California Press, Berkeley).
\bibitem[Ulmer \& Wijers 1995]{uw95} Ulmer, A., \& Wijers,
R.A.M.J. 1995, ApJ, 439, 303.
\bibitem[Weinberg 1972]{weinberg72} Weinberg, S. 1972, Gravitation and
Cosmology (Wiley, New York).
\bibitem[Wijers \& Paczy\'nski 1994]{wp94} Wijers, R.A.M.J., \&
Paczy\'nski, B. 1994, ApJ, 437, L107.

\end{thebibliography}
\end{document}